\documentclass[aip,amsmath,amssymb,reprint]{revtex4-1}

\usepackage{graphicx}
\usepackage{dcolumn}% Align table columns on decimal point
\usepackage[utf8]{inputenc}
\usepackage[T1]{fontenc}
\usepackage{mathptmx}
\usepackage{etoolbox}
\usepackage{color}

\makeatletter
\def\@email#1#2{%
 \endgroup
 \patchcmd{\titleblock@produce}
  {\frontmatter@RRAPformat}
  {\frontmatter@RRAPformat{\produce@RRAP{*#1\href{mailto:#2}{#2}}}\frontmatter@RRAPformat}
  {}{}
}%
\makeatother

\begin{document}

\title{Ab initio treatment of molecular Coster-Kronig decay using complex-scaled equation-of-motion 
coupled-cluster theory}
\author{Jan Philipp Drennhaus}
\author{Anthuan Ferino P\'erez}
\author{Florian Matz}
\author{Thomas-C. Jagau}
\email{thomas.jagau@kuleuven.be}

\affiliation{Division of Quantum Chemistry and Physical Chemistry, 
KU Leuven, Celestijnenlaan 200F, 3001 Leuven, Belgium}

\begin{abstract} \setlength{\parindent}{0pt}
Vacancies in the L$_1$ shell of atoms and molecules can decay non-radiatively via Coster-Kronig 
decay whereby the vacancy is filled by an electron from the L$_{2,3}$ shell while a second electron 
is emitted into the ionization continuum. This process is akin to Auger decay, but in contrast to Auger 
electrons, Coster-Kronig electrons have rather low kinetic energies of less than 50 eV. In the present 
work, we extend recently introduced methods for the construction of molecular Auger spectra that 
are based on complex-scaled equation-of-motion coupled-cluster theory to Coster-Kronig decay. 
We compute ionization energies as well as total and partial decay widths for the 2s$^{-1}$ states 
of argon and hydrogen sulfide and construct the L$_1$L$_{2,3}$M Coster-Kronig and L$_1$MM 
Auger spectra of these species. Whereas our final spectra are in good agreement with the available 
experimental and theoretical data, substantial disagreements are found for various branching ratios 
suggesting that spin-orbit coupling makes a major impact on Coster-Kronig decay already in the third 
period of the periodic table.
\end{abstract}

\maketitle

%%%%%%%%%%%%%%%%%%%%%%%%%%%%%%%%%%%%%%%%%%%%%%%%%%
\section{Introduction} \label{sec:intro}
Core-vacant states of atoms and molecules can relax by means of Auger decay, where the core 
vacancy is filled, while a second electron is emitted carrying away the excess energy.\cite{auger23,
agarwal13} By measuring the kinetic energy of these Auger electrons, information on the electronic 
structure of molecules,\cite{ramasesha16,norman18,kraus18} materials,\cite{hofmann12} 
surfaces,\cite{weightman82} and nanostructures\cite{raman11,unger20} can be obtained. While 
Auger electrons originating from K-shell vacancies typically have energies of hundreds or even 
thousands of electron volts, electrons originating from vacancies in higher shells can be substantially 
slower with energies in the range of 25--50 eV, i.e., more similar to intermolecular Coulombic 
decay\cite{cederbaum97,jahnke20} than to K-shell Auger decay. 

These low-energy electrons stem from Coster-Kronig transitions\cite{coster35} in which the core 
hole is filled by an electron from the same shell, while a second electron from a higher shell is 
emitted. Electronic states of atoms in the third period of the periodic table with empty 2s orbitals, 
i.e., L$_1$-shell vacancies in X-ray notation, are perhaps the simplest electronic structures where 
Coster-Kronig decay is energetically possible. These states can be specifically prepared by X-rays 
but they are also the result of Auger decay of K-shell vacancies. Because of the relatively low energy 
of Coster-Kronig electrons, they can interact more strongly with matter than Auger electrons, which 
implies that they are relevant for radiation damage of biological systems similar to other secondary 
electrons. 

In Coster-Kronig decay of L$_1$-shell vacancies, an electron from a 2p orbital, i.e., the L$_{2,3}$-shell 
in X-ray notation, fills the empty 2s orbital and an electron from the M-shell, which is formed by the 
3s, 3p, and potentially 3d orbitals, is emitted. It is well established that this process is more efficient 
than K-shell Auger decay,\cite{agarwal13,mcguire71,bambynek72,krause79a,krause79b,chen81,puri93} 
resulting in large decay widths of the order of several eV, which corresponds to extraordinarily short 
lifetimes of less than one femtosecond. 

Coster-Kronig decay of L$_1$ vacancies is always accompanied by decay channels in which both 
electrons stem from higher-lying shells, i.e., in the case of the third period of the periodic table 
L$_1$MM decay. These latter decay channels have, however, much smaller widths than the 
Coster-Kronig channels. Notably, a decay process where all three electrons stem from the 
same shell is energetically forbidden in the third period of the periodic table. These so-called 
super-Coster-Kronig transitions have, however, been described for 3d transition metals and 
heavier elements.\cite{guillot77,bruhn78,davis81} 

Ample experimental data have been reported about Coster-Kronig decay in atoms, especially 
in noble gases\cite{mehlhorn68,kylli99,lablanquie00,lablanquie11,avaldi19,boudjemia21,
hikosaka22,jureta23} but also, for example, in magnesium\cite{kochur01} and potassium.\cite{
soronen24} For argon, in particular, the Coster-Kronig spectrum and the branching ratios 
between different peaks are known with considerable precision.\cite{lablanquie11} Thanks 
to theoretical modeling using approximate Hartree-Fock theory,\cite{axelrod76} 
Dirac-Hartree-Slater theory,\cite{chen81,puri93} Dirac-Hartree-Fock theory,\cite{chen77,larkins78,
ohno84,glans93} multiconfigurational Dirac-Hartree-Fock (MCDHF) theory,\cite{bruneau83,karim84b,karim84,
karim85,liu21,boudjemia21,hikosaka22} and many-body Green's function theory \cite{ohno03a,
ohno03b} most aspects of these spectra are now well understood and signals have been 
unequivocally assigned to decay channels. 

For molecules, on the other hand, much less data are available. Noteworthy are experimental 
studies of hydrogen chloride,\cite{kaneyasu06} silicon dioxide,\cite{ramaker79} hydrogen 
sulfide,\cite{cesar90,hikosaka04} thiouracil,\cite{lever21} and solvated sodium, magnesium 
and aluminium cations.\cite{oehrwall10} \textit{Ab initio} modeling of molecular Coster-Kronig decay 
based on many-electron wave functions has not been reported, which illustrates that it is difficult 
to extend theoretical methods designed for atoms to molecules. 

Some of us recently developed a method to compute Auger decay rates from complex-scaled 
wave functions of core-ionized states.\cite{matz22,matz23a} In complex scaling (CS),\cite{aguilar71,
balslev71,moiseyev11,jagau22} the Hamiltonian has complex eigenenergies whose imaginary 
parts describe the decay width. There is thus no need to model the wave function of the emitted 
electron explicitly. Whereas the direct application of CS to the Hamiltonian only works for atoms 
but not for molecules, the approach has been extended to molecules by means of complex-scaled 
basis functions (CBFs).\cite{mccurdy78,white15a} The CBF method has been used to model K-LL 
Auger spectra in the second period of the periodic table,\cite{jayadev23,matz23b} interatomic and 
intermolecular Coulombic decay,\cite{parravicini23} autoionization of Rydberg states,\cite{creutzberg23} 
and most recently K-edge Auger decay of the zinc atom and the hexaaquazinc(II) 
complex.\cite{ferino24} 

In the present work, we extend this approach to Coster-Kronig decay of L$_1$-shell vacancies, 
taking argon and hydrogen sulfide as examples. In addition, we also study the L$_1$MM Auger 
spectra of these two species. The investigation of argon serves to establish the accuracy of our 
approach as there are two well-resolved experimental Coster-Kronig spectra available,\cite{kylli99,
lablanquie11} and in addition a theoretical spectrum based on MCDHF wave functions\cite{lablanquie11} 
as well as partial decay widths computed with MCDHF.\cite{liu21} Additional validation is provided 
by the comparison between CS and CBF results, which is only possible for atoms. With the 
investigation of hydrogen sulfide, for which only experimental spectra with much lower resolution 
are available,\cite{cesar90,hikosaka04} we show that our approach can be easily applied to molecular 
Coster-Kronig decay as well. 

The remainder of the manuscript is structured as follows: In Section \ref{sec:cd}, the details of 
our computations are given, whereas Section \ref{sec:res} presents our results for ionization 
energies and total decay widths as well as the L$_1$L$_{2,3}$M Coster-Kronig spectra and 
L$_1$MM Auger spectra of argon and hydrogen sulfide. 

%%%%%%%%%%%%%%%%%%%%%%%%%%%%%%%%%%%%%%%%%%%%%%%%%%
\section{Computational Details} \label{sec:cd}
To simulate an Auger spectrum, two quantities are needed for each decay channel: the kinetic 
energy of the Auger electrons and the decay rate. To compute these quantities, we use an 
approach that is based on complex-scaled equation-of-motion coupled-cluster (EOM-CC) 
theory.\cite{bravaya13,zuev14,white15b,white17,matz22} 

We treat Auger decay as a two-step process in which the second step, the filling of the core hole 
and the ejection of the Auger electron is independent of the creation of the core hole.\cite{wentzel27,
agren92} Because of energy conservation, the energy of the Auger electron equals the energy 
difference between the initial core-ionized state and the final doubly ionized states. To compute 
the energies of these states, we use the ionization potential and the double ionization potential 
variants of EOM-CC with singles and doubles excitations (EOM-IP-CCSD and 
EOM-DIP-CCSD).\cite{stanton93,stanton94,sattelmeyer03,bartlett09,sneskov12} 

The decay widths are obtained from EOMIP-CCSD calculations on the initial core-ionized 
states in which either the Hamiltonian is complex scaled (CS-EOMIP-CCSD) or functions with 
a complex-scaled exponent are included in the basis set (CBF-EOM-IP-CCSD). The total width 
$\Gamma$ is obtained from 
\begin{equation} \label{eq:tw}
\Gamma = 2 \cdot (\text{Im}(E_\text{coreIP}) - \text{Im}(E_0))  
\end{equation}
where $E_\text{coreIP}$ and $E_0$ are the energies of the core-ionized state and the neutral reference 
state. The optimal complex scaling angles $\theta_\text{opt}$ are determined by minimizing 
$d|(E_\text{coreIP}-E_0)|/d\theta$\cite{moiseyev78} and reported in the Supplementary Information. 

The partial widths for the decay channels are determined at $\theta_\text{opt}$ by means of the Auger 
channel projector that excludes certain amplitudes from the EOM-CC excitation manifold.\cite{matz23a} 
This can be viewed as a generalized core-valence separation.\cite{cederbaum80} Specifically, to 
compute the width $\gamma_{ij}$ of a decay channel that involves the valence orbitals $i$ and $j$, 
a complex-variable EOM-IP-CCSD calculation is performed in which the corresponding doubles 
amplitudes $r^a_{ij}$ are set to zero for all $a$. The difference between $\Gamma$ obtained in 
this calculation and $\Gamma$ from a complex-variable EOM-IP-CCSD calculation with the full 
excitation manifold defines $\gamma_{ij}$.

The Auger channel projector calculations yield the partial widths in terms of orbital pairs of the initial 
state and hence do not account for relaxation in the final states. To incorporate these relaxation effects 
into the description, the partial widths are assigned to the EOM-DIP-CCSD energies using the squared 
EOM-DIP-CCSD amplitudes as weighting factors.\cite{jayadev23,matz23b} In all calculations reported 
here, these weighting factors are close to one, indicating relatively little relaxation of the wave function 
upon filling a 2s$^{-1}$ core hole in comparison to what we observed in earlier work on 1s$^{-1}$ core 
holes. 

Notably, we were not able to describe the 2s$^{-1}$ states of argon and hydrogen sulfide in terms of 
CCSD wave functions based on core-vacant Hartree-Fock determinants. The CCSD equations for these 
states suffer from convergence problems because the unoccupied 2s orbital is too close in energy to 
other occupied orbitals. As a consequence, the evaluation of partial widths from a decomposition of 
the CCSD energy that we used previously for 1s$^{-1}$ states\cite{matz22,ferino24} is not possible 
for the 2s$^{-1}$ states that are of interest here. 

For comparison purposes, we also performed Fano-EOM-CCSD calculations in which the partial 
widths are obtained as transition amplitudes between an initial state represented by a 
core valence separation (CVS)-EOM-IP-CCSD\cite{vidal18} wave function and a 
final state represented by a product of an EOM-DIP-CCSD wave function and a plane wave.\cite{
skomorowski21a} We note that the core orbitals are frozen in the CCSD reference 
state on which the CVS-EOM-IP-CCSD calculations are based, whereas no orbitals are frozen in 
all other calculations. Additionally, we constructed spectra in which the density of EOM-DIP-CCSD 
states replaces the partial decay widths, which is equivalent to assuming that every channel has the 
same decay width. 

CS-EOM-IP-CCSD and EOM-DIP-CCSD calculations on argon were carried out using the aug-cc-pCV5Z 
basis that was further augmented by 8 complex-scaled s, p, and d-shells for the corresponding 
CBF-EOM-IP-CCSD calculations. EOM-DIP-CCSD calculations on hydrogen sulfide were done in a 
basis set denoted as aug-cc-pCVTZ(5sp), which uses s and p-shells from the aug-cc-pCV5Z basis, 
whereas the shells with higher angular momentum are taken from aug-cc-pCVTZ. For the corresponding 
CBF-EOM-IP-CCSD calculations, 4 to 8 complex-scaled s, p, and d-shells were added to the basis 
sets of sulfur and hydrogen. The exponents of all complex-scaled shells were determined using the 
procedure described in Ref. \citenum{matz22} and are reported in the Supplementary Information. They 
roughly span the range from 10 to 0.01 and include thus functions that are significantly more diffuse 
than those used in previous studies of K-shell Auger decay. 

The SH bond length and the HSH bond angle of hydrogen sulfide are 1.3338 \AA\ and 92.2$^\circ$, 
respectively, in all calculations. Core electrons were included in the correlation treatment except for 
the Fano-EOM-CCSD calculations in which the 1s and 2s orbitals were frozen. All Auger spectra are 
normalized such that the most intense peak has the same height with every computational approach. 
To construct the final spectra, we used a Lorentzian broadening function with a full width at half 
maximum (FWHM) of 2 eV, except for the L$_1$MM spectrum of argon, where the FWHM is 3 eV. 
All electronic-structure calculations were carried out using the Q-Chem program package, version 
6.0.\cite{qchem} Note that all irreducible representations are reported according to Q-Chem's convention, 
which differs from Mulliken's convention.

%%%%%%%%%%%%%%%%%%%%%%%%%%%%%%%%%%%%%%%%%%%%%%%%%%
\section{Results} \label{sec:res}
\subsection{Ionization energies} \label{sec:ie}

\begin{table} \centering
\caption{Ionization energies of the 2s$^{-1}$ states of argon and hydrogen sulfide in eV computed 
with different methods. The aug-cc-pCV5Z basis set is used for Ar, the aug-cc-pCVTZ(5sp) basis 
set for H$_2$S. In CBF calculations, the basis sets are further augmented by 8 complex-scaled 
s, p, and d-shells.}
\begin{tabular*}{\linewidth}{@{\extracolsep{\fill}} rll} \hline
 & Ar & H$_2$S \\ \hline
EOM-IP-CCSD & 326.58 & --- \\
CVS-EOM-IP-CCSD & 324.87 & 234.50 \\
CS-EOM-IP-CCSD & 325.90 & --- \\
CBF-EOM-IP-CCSD & 325.96 & 234.99 \\
Experiment\cite{glans93,hikosaka04} & 326.25 $\pm$ 0.05 & 235.0 $\pm$ 0.1 \\ \hline
\end{tabular*} \label{tab:ie}
\end{table}

Table \ref{tab:ie} shows ionization energies for the 2s$^{-1}$ states of argon and hydrogen sulfide 
computed with different flavors of EOM-IP-CCSD. The corresponding double ionization energies 
are reported in the Supplementary Information. The CBF-EOM-IP-CCSD results in Table \ref{tab:ie} 
agree with the experimentally determined ionization energies for argon\cite{glans93} and H$_2$S\cite{
hikosaka04} within 0.3 eV and less than 0.1 eV, respectively, although we note that a rigorous 
comparison would require the consideration of triple excitations as well as relativistic corrections, 
and for H$_2$S also the treatment of vibrational effects. 

Our results for argon illustrate that complex scaling decreases the ionization energy 
by about 0.7 eV even though a very large basis set is used. The difference between complex scaling 
of the Hamiltonian and of the basis set is, however, negligible. Notably, CVS decreases 
the energy by 1.7 eV with respect to regular EOM-IP-CCSD and by 1.0 eV with respect to 
CBF-EOM-IP-CCSD, leading to a significantly less good agreement with the experiment. This is 
similar to the 1.3 eV difference between CBF-EOM-IP-CCSD and CVS-EOM-IP-CCSD that we 
observed in recent work on 1s$^{-1}$ states of benzene.\cite{jayadev23} In the case of H$_2$S, 
the difference between CBF-EOM-IP-CCSD and CVS-EOM-IP-CCSD amounts to only 0.5 eV. 
Here, we were, however, not able to converge the EOM-IP-CCSD equations without CVS or 
employing CBFs. 

%%%%%%%%%%%%%%%%%%%%%%%%%%%%%%%%%%%%%%%%%%%%%%%%%%
\subsection{Decay width of the 2s$^{-1}$ state of argon} \label{sec:tw1}

\begin{table*}[tbh] \centering
\caption{Total decay widths and sum of partial decay widths of the 2s$^{-1}$ states of argon and 
hydrogen sulfide in meV computed with different methods. For CBF calculations, the complex-scaled 
shells are denoted in italics. Experimental values are given as well.}
\begin{tabular*}{\linewidth}{@{\extracolsep{\fill}} rrrrrr} \hline
Method & Basis set & Total width & Sum of all & \multicolumn{2}{c}{Sum of partial widths of} \\ 
 &  & from Eq. \eqref{eq:tw} & partial widths & L$_{2,3}$M channels & MM channels \\ \hline
\multicolumn{6}{c}{Argon} \\ \hline
CS-EOM-CCSD & aug-cc-pCVQZ & 2531.8 & 2347.3 & 2273.2 & 74.1 \\
CS-EOM-CCSD & aug-cc-pCV5Z & 2632.3 & 2450.2 & 2373.3 & 77.0 \\
CBF-EOM-CCSD & aug-cc-pCV5Z+\textit{4(spd)} & 1053.8 & 872.2 & 820.0 & 52.4 \\
CBF-EOM-CCSD & aug-cc-pCV5Z+\textit{6(spd)} & 2100.9 & 2294.6 & 2216.4 & 78.2 \\
CBF-EOM-CCSD & aug-cc-pCV5Z+\textit{8(spd)} & 2668.6 & 2334.2 & 2259.0 & 75.3 \\
Semi-empirical theory\cite{krause79a} &  & \multicolumn{2}{c}{1630} &  &  \\ 
Dirac-Hartree-Slater\cite{chen81} &  &  \multicolumn{2}{c}{2716} & 2595 & 121 \\
Dirac-Hartree-Fock\cite{glans93} &  & \multicolumn{2}{c}{1850} &  &  \\
MCDHF\cite{kylli99} &  &  &  & 2330 & \\
MCDHF\cite{liu21} &  & \multicolumn{2}{c}{2092} & 2037 & 55 \\
Experiment\cite{glans93} &  & \multicolumn{2}{c}{2250 $\pm$ 50} &  &  \\
Experiment\cite{mehlhorn68} &  & \multicolumn{2}{c}{1840 $\pm$ 200} &  &  \\ \hline
\multicolumn{6}{c}{Hydrogen sulfide} \\ \hline
CBF-EOM-CCSD & aug-cc-pCVTZ(5sp)+\textit{4(spd)} & 1119.1 & 1020.0 & 963.4 & 56.9 \\
CBF-EOM-CCSD & aug-cc-pCVTZ(5sp)+\textit{6(spd)} & 1603.2 & 1407.4 & 1362.4 & 44.9 \\
CBF-EOM-CCSD & aug-cc-pCVTZ(5sp)+\textit{8(spd)} & 1672.2 & 1440.5 & 1396.5 & 44.1 \\ 
Semi-empirical theory\cite{krause79a} & \multicolumn{1}{l}{(S atom)} & 1490 & \\ 
Approximate HF theory \cite{axelrod76} & \multicolumn{1}{l}{(S atom)} & 2590 & \\
Experiment\cite{hikosaka04} & & 1800 & \\ \hline
\end{tabular*} \label{tab:tw}
\end{table*}

The upper part of Table \ref{tab:tw} shows total decay widths for the 2s$^{-1}$ state of argon 
computed with CS-EOM-IP-CCSD and CBF-EOM-IP-CCSD using different basis sets. It is seen 
that this state has a decay width of more than 2 eV, which corresponds to a very short lifetime 
of less than one third of a femtosecond. The width is 4--5 times as large as that of the 1s$^{-1}$ 
state of argon (0.46 eV)\cite{vayrynen83} and almost 10 times as large as that of the 1s$^{-1}$ 
state of neon (0.26 eV).\cite{mueller17}

If double Auger decay and other processes involving more than two electrons are 
neglected, the 2s$^{-1}$ state of argon has 14 decay channels that are in principle open, meaning 
that the final electronic state has a lower energy so that the decay process is energetically allowed. 
The partial widths for these 14 channels are reported in the SI. 8 of them involve the 2p shell and 
form the L$_1$L$_{2,3}$M Coster-Kronig spectrum. These 8 channels account for more than 96\% 
of the total decay width, whereas the remaining 6 channels, which form the L$_1$MM Auger spectrum, 
account for less than 4\%. A conspicuous difference to K-shell Auger decay is the 25\% contribution 
that the triplet decay channels deliver to the total width. By contrast, triplet states contribute only 
6\% to the decay width of the 1s$^{-1}$ state of neon.\cite{albiez90,mueller17} Notably, MCDHF 
calculations, which take account of spin-orbit coupling, yielded a 55\% contribution of triplet channels 
to the width of the 2s$^{-1}$ state of argon.\cite{liu21}

The comparison of our results to the experimental value of 2.25 eV\cite{glans93} suggests that 
the sum of partial widths is a better estimate of the total width than the value obtained from Eq. 
\eqref{eq:tw}. Using the former approach, CBF-EOM-IP-CCSD overestimates the experimental 
values by less than 4\%. Interestingly, Dirac-Hartree-Fock theory combined with the Green's 
function method yielded a value for the total width that is 20\% lower, while MCDHF theory yielded 
a value that is only 10\% lower.\cite{liu21} This suggests that electron correlation increases the 
decay width, which is in line with previous results for other electronic resonances.\cite{jagau22} 
We also note a second MCDHF value\cite{kylli99} for the sum of the width of the L$_{2,3}$M 
Coster-Kronig channels that differs from our result by no more than 3\%. 

The significant difference of 7--20\% between the sum of all partial widths and the total width evaluated 
according to Eq. \eqref{eq:tw} is similar to what has been observed in previous treatments of Auger 
decay with CBF- and CS-EOM-CCSD. It can be traced back to EOM-IP-CCSD doubles amplitudes 
$r_{ij}^a$ where $i$ or $j$ is a core orbital. The resulting configurations in the EOM-IP-CCSD wave 
function where the 2s$^{-1}$ orbital is unoccupied do not correspond to open decay channels as 
they are too high in energy, but they deliver a non-zero contribution to the total width. This effect 
is, in principle, present in every CS or CBF calculation, but it is in the present case apparently more 
pronounced in the CBF calculations. Table \ref{tab:tw} shows deviations of 330 meV (14\%) between 
$\Gamma$ from Eq. \eqref{eq:tw} and the sum of partial widths for CBF-EOM-IP-CCSD in the largest 
basis set, whereas this value amounts to 180 meV (7\%) for CS-EOM-IP-CCSD. Between each other, 
CS and CBF calculations differ by no more than 5\%.

Table \ref{tab:tw} also illustrates a need for large basis sets, which is typical of complex-scaled 
calculations. However, there are some aspects that are different from calculations on 1s$^{-1}$ states: 
Firstly, the aug-cc-pCVQZ basis already recovers 96\% of the total decay width in the present case, 
whereas this value amounted to only 64\% in previous CS-EOM-IP-CCSD calculations on the 1s$^{-1}$ 
state of neon.\cite{matz22} Secondly, more diffuse complex-scaled shells are required for the description 
of Coster-Kronig decay than for the description of K-shell Auger decay. Whereas 2--3 complex-scaled 
s, p, and d shells are sufficient for K-shell Auger decay, Table \ref{tab:tw} demonstrates that the use of 
4 complex-scaled s, p, and d-shells produces a width for the 2s$^{-1}$ state of argon that is too small 
by a factor of ca. 2.7. Upon including 6 complex-scaled shells the sum of partial widths is recovered 
almost in full, but the branching ratios between the channels still change substantially if two further 
shells are added as is apparent from the values reported in the SI. This need for more diffuse shells 
may be related to the substantially lower energy of the emitted electron in Coster-Kronig decay as 
compared to K-shell Auger decay.\cite{matz22}

Notably, the basis-set dependence of the decay channels is very different. The 6 MM decay channels 
as well as some of the 8 L$_{2,3}$M Coster-Kronig channels are already well described with 6 or even 
4 complex-scaled s, p, and d shells, whereas the width of other channels changes by more than a factor 
of 3 when going from 6 to 8 complex-scaled shells. Also, we note that the agreement between CBF- 
and CS-EOM-IP-CCSD is somewhat better for the MM decay channels than for the L$_{2,3}$M channels.

%%%%%%%%%%%%%%%%%%%%%%%%%%%%%%%%%%%%%%%%%%%%%%%%%%%

\subsection{Decay width of the 2a$_1^{-1}$ state of hydrogen sulfide} \label{sec:tw2}
The lower part of Table \ref{tab:tw} shows total decay widths for the 2a$_1^{-1}$ state of H$_2$S. 
Because of the lower point group, the 14 decay channels of the 2s$^{-1}$ state of argon correspond 
to 40 channels in the case of H$_2$S. Partial widths for all of them are reported in the SI. There are 
24 L$_{2,3}$M channels, which form the Coster-Kronig spectrum and account for 97\% of the total 
decay width, while the remaining 16 MM channels account for only 3\% of the width. Similar to argon, 
triplet channels contribute ca. 25\% to the decay width, both for the L$_{2,3}$M and the MM channels. 
Interestingly, Fano-EOM-CCSD yields a triplet contribution of 82\%, which neither agrees with 
CBF-EOM-IP-CCSD nor MCDHF results for argon. Notably, it has been argued that the representation 
of the emitted electron by a plane wave in Fano-EOM-CCSD calculations may lead to an overestimation 
of the triplet contribution.\cite{skomorowski21b} 

Whereas CBF-EOM-IP-CCSD yields very similar branching ratios for argon and H$_2$S, 
a big difference is found for the total widths themselves as that of the 2a$_1^{-1}$ state of H$_2$S is 
only 62\% of that of the 2s$^{-1}$ state of argon. Very similar ratios are observed for the widths of the 
L$_{2,3}$M and MM channels separately. Although this is qualitatively in line with previous results that 
found a stronger dependence on nuclear charge for Coster-Kronig widths than for K-shell Auger widths, 
the comparison of the experimentally determined widths of argon (2.25 eV)\cite{glans93} and hydrogen 
sulfide (1.80 eV)\cite{hikosaka04} delivers a value of 80\% for this ratio. 

Similar to argon, we observe a significant difference of ca. 15\% between the sum of all partial widths 
and the total width from Eq. \eqref{eq:tw}. Different from argon, however, the value from Eq. \eqref{eq:tw} 
is in better agreement with the experiment. A rigorous statement about the exact value of the total 
width is difficult to make because only one experimental value and no other theoretical values have 
been reported for H$_2$S. Also, there is a big disagreement of more than 1 eV between values 
computed for the total width of the sulfur atom with lower-level theories.\cite{axelrod76,krause79a} 
In any case, the 2a$_1^{-1}$ state of H$_2$S is much broader than the 1a$_1^{-1}$ states of 
H$_2$S (0.59 eV)\cite{krause79a} and H$_2$O (0.16 eV),\cite{sankari03} illustrating again the 
efficiency of Coster-Kronig decay.

We note that the basis-set dependence of the width is somewhat less pronounced for H$_2$S than for 
argon. With 6 complex-scaled s, p, and d shells, more than 95\% of the total width are captured and 
almost all partial widths are converged as well. This may be related to the lower point group of H$_2$S 
as similar trends were observed for K-shell Auger decay before. Interestingly, there are 3 decay channels 
of H$_2$S that have negative widths of 5 to 15 meV even in the largest basis set. This unphysical result 
has not been encountered for K-shell Auger decay and may indicate incompleteness of the basis set. 

%%%%%%%%%%%%%%%%%%%%%%%%%%%%%%%%%%%%%%%%%%%%%%%%%%%

\subsection{L$_1$L$_{2,3}$M Coster-Kronig spectrum of argon} \label{sec:arck}

\begin{figure}[hbt] \centering
\includegraphics[scale=0.60]{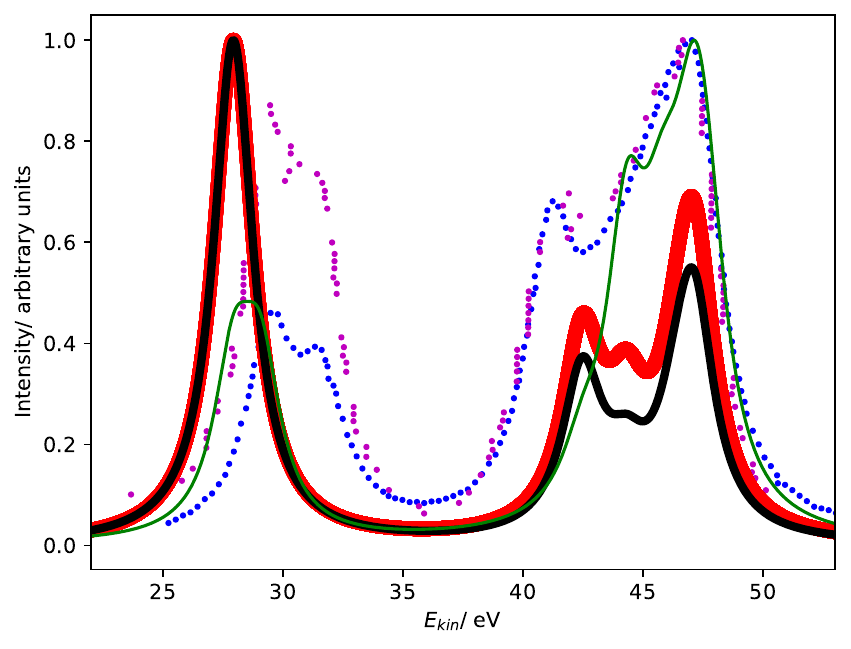}
\caption{L$_1$L$_{2,3}$M Coster-Kronig spectrum of argon. Partial decay widths were computed with 
CS-EOM-CCSD (red solid line) and CBF-EOM-CCSD (black solid line), and assuming the same width 
for every channel (green solid line). The experimental Coster-Kronig spectra reported in Refs. \citenum{kylli99} 
and \citenum{lablanquie11} are shown as blue and purple dotted lines, respectively. The theoretical 
spectra are shifted to higher kinetic energy by 3.4 eV.}
\label{fig:ar-1a}
\end{figure}
% Lorentzian broadening with FWHM = 2~eV.

\begin{figure}[hbt] \centering
\includegraphics[scale=0.60]{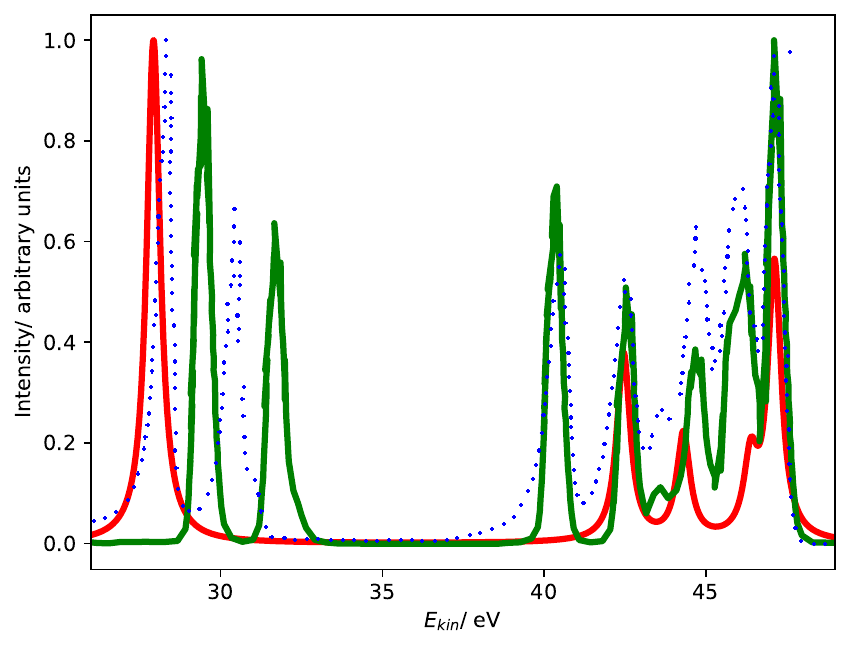}
\caption{L$_1$L$_{2,3}$M Coster-Kronig spectrum of argon. Comparison of CS-EOM-CCSD results 
(red solid line, this work) with MCDHF results (green solid line, Ref. \citenum{lablanquie11}) and the 
experimental spectrum (blue dotted line, Ref. \citenum{lablanquie11}) The theoretical spectra are 
shifted to higher kinetic energy by 3.4 eV and 1.3 eV, respectively.}
\label{fig:ar-1b}
\end{figure}
% Lorentzian broadening with FWHM = 0.5~eV.

Figure \ref{fig:ar-1a} compares the Coster-Kronig spectra of argon computed with CS- and 
CBF-EOM-IP-CCSD to two experimental spectra.\cite{kylli99,lablanquie11} In addition, we 
compare in Figure \ref{fig:ar-1b} our CS-EOM-IP-CCSD spectrum to results from MCDHF 
calculations and experimental data that were obtained from Auger multi-electron coincidence 
spectroscopy.\cite{lablanquie11} Because the resolution of the experimental data is higher, we 
applied a broadening function with a FWHM of only 0.5 eV to the theoretical spectra in this figure. 
In addition, the convergence of the CBF-EOMIP-CCSD spectrum with respect to the number of 
complex-scaled shells in the basis set is illustrated in the Supplementary Information. 

It is seen from Figure \ref{fig:ar-1a} that the Coster-Kronig spectrum of argon consists of 
two features at 27--33 eV and 38--48 eV, which correspond to the L$_{2,3}$M$_1$ and 
L$_{2,3}$M$_{2,3}$ channels, respectively. While the intensity is evenly split between 
these two features in our computations, the experiments found an intensity distribution of 
23:77\cite{kylli99} and 27:73,\cite{lablanquie11} respectively, in favor of the L$_{2,3}$M$_{2,3}$ 
channels, and MCDHF calculations delivered a ratio of 33:67.\cite{liu21} Interestingly, assuming 
that every channel has the same width delivers a ratio of 25:75, in good agreement with the 
experiment. 

Furthermore, Figures \ref{fig:ar-1a} and \ref{fig:ar-1b} show that the experimental spectra and 
the theoretical MCDHF spectrum have two peaks in the L$_{2,3}$M$_1$ region below 33 eV, 
whereas our spectra have just one peak. This mismatch is related to the $^3$P (2p$^{-1}$3s$^{-1}$) 
state having zero intensity in our computations but accounting for 250 meV in the MCDHF 
computations.\cite{liu21} Also in the experiment, the $^1$P and $^3$P states are both clearly 
visible. Similar disagreements are also present in the L$_{2,3}$M$_{2,3}$ region in Figure 
\ref{fig:ar-1b}: Our calculations yield 4 peaks each corresponding to one decay channel, 
whereas MCDHF yields 6 peaks some of which are composed of more than one channel. 

All of these shortcomings suggest that spin-orbit coupling, which is missing in our theoretical model, 
changes the intensity distribution in the Coster-Kronig spectrum of argon significantly. Notably, the 
importance of spin-orbit interaction for the branching ratio between the $^1$P (2p$^{-1}$3s$^{-1}$) 
and $^3$P (2p$^{-1}$3s$^{-1}$) states was established already 40 years ago.\cite{bruneau83,
karim84,karim85} However, it should also be noted that the overall shape of the experimental 
spectrum is well reproduced by our computations despite the neglect of spin-orbit coupling. 

%%%%%%%%%%%%%%%%%%%%%%%%%%%%%%%%%%%%%%%%%%%%%%%%%%

\subsection{L$_1$L$_{2,3}$M Coster-Kronig spectrum of hydrogen sulfide} \label{sec:h2sck}

\begin{figure} \centering
\includegraphics[scale=0.60]{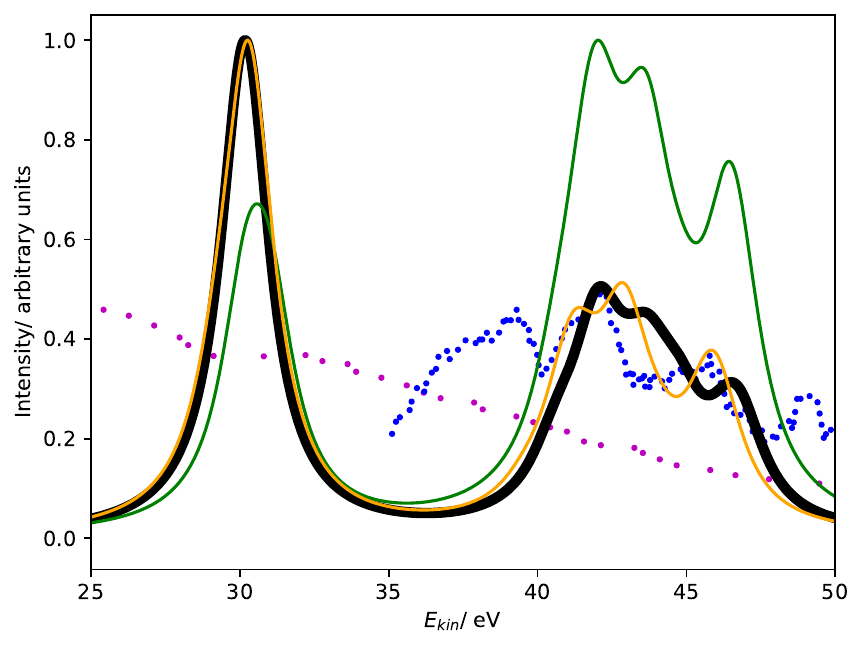}
\caption{L$_1$L$_{2,3}$M Coster-Kronig spectrum of hydrogen sulfide. Partial decay widths were 
computed with CBF-EOM-CCSD (black solid line), Fano-EOM-CCSD (orange solid line), and 
assuming the same width for every channel (green solid line). The experimental data reported 
in Refs. \citenum{hikosaka04} and \citenum{cesar90} are shown as blue and purple dotted lines. 
The theoretical spectra are shifted to higher kinetic energy by 7.5 eV.}
\label{fig:h2s-1}
\end{figure}
% Lorentzian broadening with FWHM = 2~eV.

Figure \ref{fig:h2s-1} shows the Coster-Kronig spectra of hydrogen sulfide computed with 
CBF-EOM-IP-CCSD and Fano-EOM-CCSD as well as the available experimental results, 
which are of lower quality than in the case of argon. The convergence of the CBF-EOM-IP-CCSD 
spectrum with respect to the number of complex-scaled shells in the basis set is illustrated in the 
Supplementary Information. 

As expected, this spectrum has the same general structure as that of argon shown in Figure 
\ref{fig:ar-1a}. It consists of two features corresponding to the L$_{2,3}$M$_1$ and L$_{2,3}$M$_{2,3}$ 
decay channels. Similar to argon, CBF-EOM-IP-CCSD delivers a roughly even distribution of the 
intensity between the two features. The feature at lower energy is composed of 3 singlet decay 
channels involving the 4a$_1$ orbital, which forms the M$_1$ shell, and the 3a$_1$, 1b$_1$, 
and 1b$_2$ orbitals, which form the L$_{2,3}$ shell. Notably, the 3 corresponding triplet channels 
have a slightly negative decay width in the CBF-EOM-IP-CCSD calculations, indicating basis-set 
incompleteness. The feature at higher energy is composed of 18 L$_{2,3}$M$_{2,3}$ decay 
channels, where M$_{2,3}$ = 2b$_1$, 5a$_1$, 2b$_2$. 

In the Fano-EOM-CCSD treatment, the intensity is also roughly evenly distributed between the 
L$_{2,3}$M$_1$ and L$_{2,3}$M$_{2,3}$ features. Interestingly, the L$_{2,3}$M$_1$ triplet 
channels that have zero intensity in the CBF-EOM-IP-CCSD calculations are very pronounced 
with Fano-EOM-CCSD. The final spectra computed with the two methods are, however, in fairly 
good agreement.

Given the disagreements we observed for argon between CBF-EOM-IP-CCSD on the one hand 
and MCDHF and the experimental data on the other hand, the correctness of the H$_2$S 
spectra in Figure \ref{fig:h2s-1} and the branching ratios can be questioned. Unfortunately, 
the experimental H$_2$S spectrum\cite{hikosaka04} only covers the energy range between 
35 and 50 eV, i.e., it does not cover the L$_{2,3}$M$_1$ feature so that a definitive statement 
is difficult. We note, however, that theory and experiment agree about the L$_{2,3}$M$_{2,3}$ 
feature of the spectrum having a different shape for H$_2$S than for argon. 

Regarding the Auger electron energies, our calculations suggest that the two features of the 
Coster-Kronig spectrum lie somewhat closer to each other in the case of H$_2$S as compared 
to argon. Whereas the L$_{2,3}$M$_1$ feature is only 1 eV lower in energy for hydrogen sulfide 
than for argon, the L$_{2,3}$M$_{2,3}$ feature moves by 4--5 eV. Note that the trends in the 
absolute energies are not immediately apparent from Figures \ref{fig:ar-1a} and \ref{fig:h2s-1} 
as different shifts were applied to the theoretical spectra. Also, because the experimental 
spectrum for H$_2$S is incomplete, it cannot be confirmed if the trend is correct. 

%%%%%%%%%%%%%%%%%%%%%%%%%%%%%%%%%%%%%%%%%%%%%%%%%%

\subsection{L$_1$MM Auger spectrum of argon} \label{sec:arlmm}

\begin{figure} \centering
\includegraphics[scale=0.60]{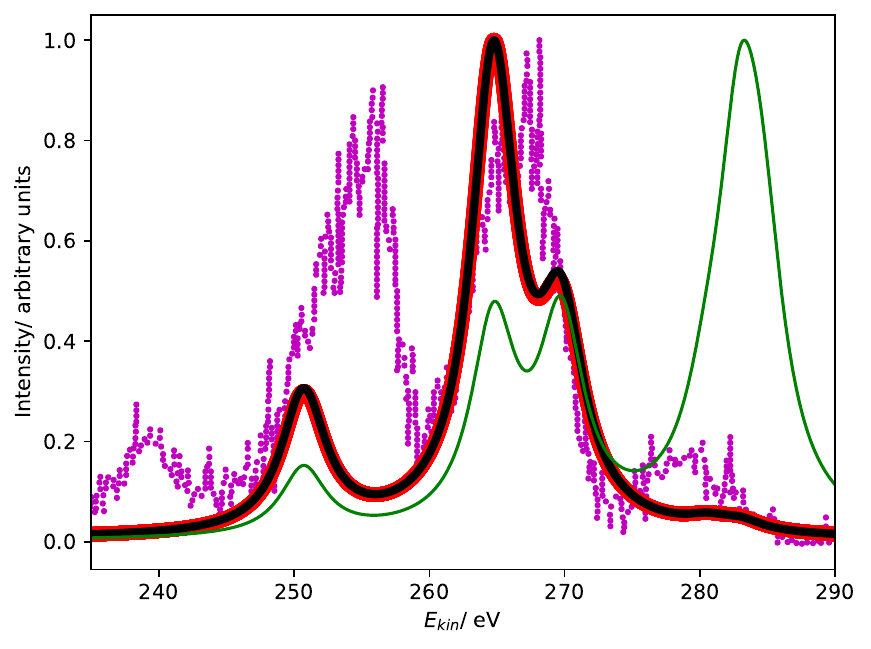}
\caption{L$_1$MM Auger spectrum of argon. Partial decay widths were computed with 
CS-EOM-CCSD (red solid line), CBF-EOM-CCSD (black solid line), and assuming the 
same width for every channel (green solid line). The experimental spectrum reported in 
Ref. \citenum{lablanquie11} is shown as purple dotted line. The theoretical spectra are shifted 
to higher kinetic energy by 2.0 eV.}
\label{fig:ar-2}
\end{figure}

Figure \ref{fig:ar-2} compares the L$_1$MM Auger spectra computed with CS-EOM-IP-CCSD 
and CBF-EOM-IP-CCSD to the experimental spectrum.\cite{lablanquie11} The theoretical spectra 
consist of two features: The feature at an Auger electron energy of around 250 eV corresponds 
to the M$_1$M$_1$ (3s$^{-2}$) channels, whereas the broader feature with two peaks between 
262 eV and 272 eV corresponds to the M$_1$M$_{2,3}$ (3s$^{-1}$3p$^{-1}$) channels. Notably, 
the M$_{2,3}$M$_{2,3}$ (3p$^{-2}$) channels have very low intensity in our calculations and are 
barely visible in Figure \ref{fig:ar-2}. 

Despite the fairly low resolution of the experimental spectrum, which is a consequence of the low 
intensity of the MM channels, a mismatch with the theoretical spectrum about the distribution of 
intensity between the M$_1$M$_1$ and M$_1$M$_{2,3}$ channels is apparent: In the theoretical 
spectra, the M$_1$M$_{2,3}$ channels account for 80\% of intensity, whereas a roughly even 
distribution is found in the experiment. The low intensity of the M$_{2,3}$M$_{2,3}$ channels is, 
however, found in the experiment as well. 

%%%%%%%%%%%%%%%%%%%%%%%%%%%%%%%%%%%%%%%%%%%%%%%%%%

\subsection{L$_1$MM Auger spectrum of hydrogen sulfide} \label{sec:h2slmm}

\begin{figure} \centering
\includegraphics[scale=0.60]{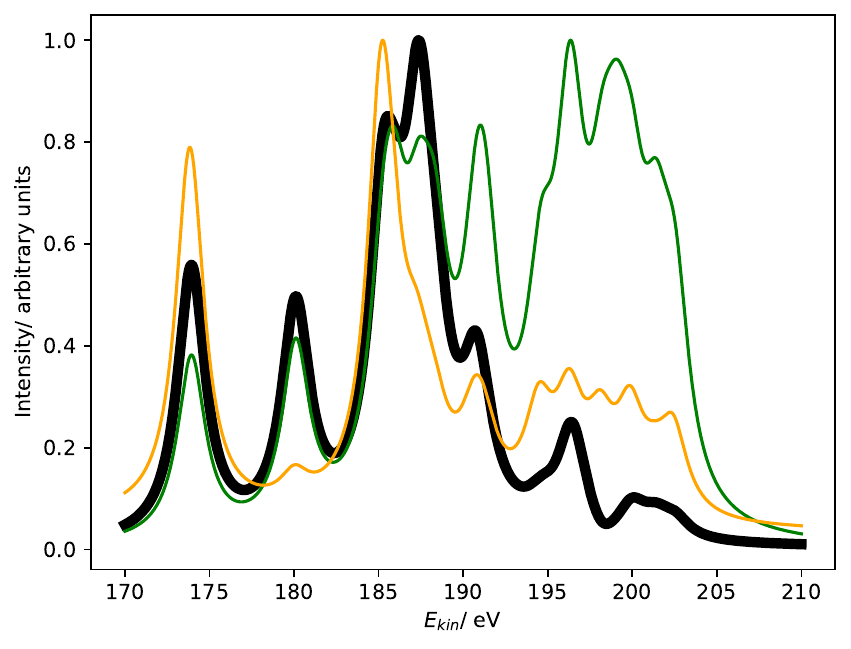}
\caption{L$_1$MM Auger spectrum of hydrogen sulfide. Partial decay widths were computed with 
CBF-EOM-CCSD (black solid line), Fano-EOM-CCSD (orange solid line), and assuming the same 
width for every channel (green solid line).}
\label{fig:h2s-2}
\end{figure}

Figure \ref{fig:h2s-2} shows the L$_1$MM Auger spectra of hydrogen sulfide computed with 
CBF-EOM-IP-CCSD and Fano-EOM-CCSD. Although there is no experimental spectrum available, 
several differences between this spectrum and the corresponding spectrum of argon in Figure 
\ref{fig:ar-2} are interesting. First, the spectrum covers a different energy range extending roughly 
from 170 to 205 eV, whereas the L$_1$MM spectrum of argon extends from 245 to 280 eV. This 
is a direct consequence of the MM double ionization energies differing by no more 5 eV between 
argon and H$_2$S, while the core ionization energies (see Table \ref{tab:ie}) differ by 80 eV. 
Notably, the Coster-Kronig spectra shown in Figures \ref{fig:ar-1a} and \ref{fig:h2s-1} cover a 
very similar energy range because the energies of initial and final states are subject to almost 
the same shift when going from argon to H$_2$S.

Second, the L$_1$MM Auger spectrum of H$_2$S computed with CBF-EOM-IP-CCSD has a 
different structure than that of argon comprising 7 peaks as compared to 3. This is again different 
to the Coster-Kronig spectrum, where the differences between argon and hydrogen sulfide are 
more subtle. The first peak from the left in Figure \ref{fig:h2s-2} at around 173 eV corresponds 
to the M$_1$M$_1$ (4a$_1^{-2}$) channel, whereas the second peak at 180 eV and the feature 
between 183 and 191 eV stem from the M$_1$M$_{2,3}$ channels. The 2 remaining peaks at 
196 eV and 200 eV correspond to the M$_{2,3}$M$_{2,3}$ channels, which account for a 
contribution of 10\% to the total L$_1$MM width in H$_2$S as opposed to a negligible contribution 
in argon. Interestingly, the branching ratio between the M$_1$M$_1$ and the M$_1$M$_{2,3}$ 
channels only changes from 80:20 to 72:17 when going from argon to hydrogen sulfide.

%%%%%%%%%%%%%%%%%%%%%%%%%%%%%%%%%%%%%%%%%%%%%%%%%
\section{Conclusions} \label{sec:conc}
We have investigated the nonradiative decay of the 2s$^{-1}$ states of argon and hydrogen 
sulfide using the EOM-IP-CCSD method combined with complex scaling of the Hamiltonian 
or, alternatively, the basis set. These 2s$^{-1}$ states have lifetimes of less than 1 femtosecond 
and are thus much shorter-lived than 1s$^{-1}$ states of light elements, which reflects the 
efficiency of L$_1$L$_{2,3}$M Coster-Kronig decay whereby an L$_1$-core hole is filled by 
an electron from the L$_{2,3}$-shell. 

In agreement with previous investigations, we find that Coster-Kronig decay channels account 
for more than 95\% of the total decay width of 2s$^{-1}$ states. This branching ratio is very 
similar for argon and H$_2$S, but the total width of 2s$^{-1}$ states depends more strongly 
on the nuclear charge than that of 1s$^{-1}$ states. Theory and experiment agree about these 
trends qualitatively, but there remain several discrepancies about other trends. Firstly, according 
to our CBF-EOM-IP-CCSD results, the 2s$^{-1}$ state of H$_2$S has 62\% of the width of the 
corresponding state of argon, while experiment suggests a ratio of 80\%. Secondly, 
CBF-EOM-IP-CCSD suggests for argon and hydrogen sulfide a contribution of 25\% by triplet 
decay channels, whereas Fano-EOM-CCSD yields a contribution of more than 80\% for H$_2$S 
and previous MCDHF calculations yielded a value of 55\% for argon. All in all, however, there 
can be no doubt that triplet decay channels are more important for L$_1$L$_{2,3}$M 
Coster-Kronig decay than for KLL Auger decay. A third discrepancy occurs for the 
L$_1$L$_{2,3}$M$_1$:L$_1$L$_{2,3}$M$_{2,3}$ branching ratio where CBF-EOM-IP-CCSD 
suggests equal contributions, whereas the L$_1$L$_{2,3}$M$_{2,3}$ channels account for 
ca. 75\% of intensity in experiments on argon and MCDHF calculations suggest a contribution 
of 67\%.

Despite these substantial discrepancies, the final L$_1$L$_{2,3}$M Coster-Kronig spectra 
and L$_1$MM Auger spectra obtained with different theoretical methods are in fairly good 
agreement with each other and also with the available experimental data. Notably, 
Coster-Kronig electrons emitted by argon and H$_2$S have approximately the same energy, 
whereas electrons stemming from L$_1$MM Auger decay are about 80 eV faster for argon. 
Also, the L$_1$L$_{2,3}$M Coster-Kronig spectra differ much less between the two species 
than the L$_1$MM Auger spectra. 

Besides these results on 2s$^{-1}$ states, our work offers insights into the workings of the 
method of complex basis functions: Because of the simultaneous presence of L$_1$L$_{2,3}$M 
decay channels that produce electrons with kinetic energies of only 25 to 50 eV and L$_1$MM 
decay channels that produce electrons with kinetic energies of more than 100 eV, steep and diffuse 
complex-scaled basis functions are required at the same time. As a result, larger basis sets are 
needed for the description of 2s$^{-1}$ states with the CBF method than for the description of 
1s$^{-1}$ states. 

Our results, in particular the discrepancies between different theoretical approaches, also 
demonstrate the need for further experimental and theoretical work in the area of Coster-Kronig 
decay, especially about molecules. We believe that the CBF method offers some critical 
advantages for such theoretical investigations: Foremost, atoms and molecules can be treated 
on an equal footing at the same level of accuracy. Also, the total width can be accessed more 
easily than with approaches that rely on a channel-by-channel treatment. At the same time, our 
work illustrates the need for further development: In particular, the consideration of spin-orbit 
coupling in CBF-EOM-CC and Fano-EOM-CC calculations is likely to change several branching 
ratios significantly. 

%%%%%%%%%%%%%%%%%%%%%%%%%%%%%%%%%%%%%%%%%%%%%%%%%
\section*{Conflicts of interest}
There are no conflicts to declare.

\section*{Acknowledgements}
T.-C. J. gratefully acknowledges funding from the European Research Council (ERC) under the 
European Union's Horizon 2020 research and innovation program (Grant Agreement No. 851766) 
and the KU Leuven internal funds (Grant No. C14/22/083).

\bibliography{ck} 

\end{document}